\documentclass[a4paper,12pt]{article}

\usepackage{epsfig}
\usepackage[cp 1251]{inputenc}

\begin{document}

\begin{center}
{\Large\bf Search for the $\Delta^{++}$ component in the $^{12}$C ground
state using the $^{12}$C$(\gamma,\pi^{+}p)$ reaction }\\

\vspace{3mm}
{\large V.M. Bystritsky$^a$, A.I. Fix$^b$, I.V. Glavanakov$^c$,
P. Grabmayr$^d$, \\ Yu.F. Krechetov$^c$\footnote{
Nuclear Physics Institute at Tomsk Polytechnic University, P.O.
Box 25, 634050 Tomsk, Russia. \mbox{Tel. +7 3822 423994},
email: krechet@npi.tpu.ru}, O.K. Saigushkin$^c$,
E.N. Schuvalov$^c$, \\ A.N. Tabachenko$^c$}\\

{\small\it $^a$ Joint Institute for Nuclear Research, 141980 Dubna,
Russia\\
$^b$ Tomsk Polytechnic University, 634034 Tomsk, Russia\\
$^c$ Nuclear Physics Institute
at Tomsk Polytechnic University, 634050 Tomsk, Russia\\
$^d$ Physikalisches Institut, Universit\"at T\"ubingen,
D-72076 T\"ubingen, Germany}
\end{center}

\begin{abstract}
The differential cross section for the $^{12}$C$(\gamma,\pi^{+}p)$ reaction
has been measured in the $\Delta$(1232) resonance region at high recoil
momenta of the residual nuclear system.  The data are analyzed under the
assumption that the formation of the $\pi^+p$ pairs may be interpreted as a
$\gamma\Delta^{++}\rightarrow\pi^+p$ process which takes place on a
$\Delta^{++}$ preexisting in the target nucleus.
Estimates of the $\Delta^{++}$ momentum distribution 
$\rho_{\Delta^{++}}(\overline{p})$ = 0.17~fm$^3$ for a mean momentum
$\overline{p}=300\pm49$~MeV/c as well as the number of $\Delta$ isobars 
per nucleon $N_\Delta=0.017$ were obtained for $^{12}$C.
\vspace*{3mm}

\end{abstract}

\section{Introduction}

The interaction of energetic projectiles with nuclei at high momentum transfer
implies probing the nuclear structure at medium and at short internucleon
distances. In short range nucleon-nucleon dynamics a wide variety of
developments concerns the internal nucleon degrees of freedom. The nontrivial
substructure of a nucleon manifests itself in the existence of the internally
excited states, i.e.\ baryon resonances or isobars. One possibility to
describe these subnucleonic effects in modern nuclear models is the addition
of exotic components, the so-called isobar configurations, to the conventional
nuclear wave functions.

Comparing different approaches (for comprehensive reviews see
e.g.~\cite{Green76,Aren78,Anastas}) it was found that the strongest isobar
admixture in nuclei stems from the $\Delta$(1232) resonance. The exchange of
$\pi$ and $\rho$ mesons may result in virtual $N\Delta$ and $\Delta\Delta$
states.  Most theoretical and experimental investigations of the $\Delta$
isobar configuration have been done for the few-body systems $^{2,3}$H and
$^3$He, and to a smaller degree for heavier nuclei. At the same time, the
amount of the virtual $\Delta$ isobars in heavier nuclei with their higher
density is expected to be more essential than in lighter nuclei. Here we would
like to mention the explicit calculation~\cite{Horl78} which resulted in a
3.2\% probability for $\Delta$ components in the $^{12}$C ground state.

As for the experimental search for the isobar admixture in nuclei with
$A\geq4$, the observation of $\Delta$ knock-out from $^9$Be~\cite{ABC94} and
the recent double charge exchange experiment $(\pi^+,\pi^-)$ on $^3$H, $^4$He,
$^6$Li, $^7$Li, $^{12,13}$C, $^{90}$Zr, and $^{208}$Pb~\cite{Morr98} are of
the greatest interest. The first experiment~\cite{ABC94} was carried out at
the proton beam of the synchrocyclotron at the Petersburg Nuclear Physics
Institute where the inclusive differential cross sections for the
$^9$Be($p,^8$He) reaction at the proton energy of 1~GeV have been measured. An
enhancement of the $^8$He yield was observed for the spectator momentum less
than 300~MeV/c, which was connected with the $\Delta^{++}$ knock-out from
$^9$Be.  The second experiment~\cite{Morr98} was performed using the 500~MeV
pion beam at the Clinton P. Anderson Meson Physics Facility. The cross
sections for the $(\pi^+,\pi^{\pm}p)$ reaction have been measured in
quasi-free kinematics. The observed enhancement of the $(\pi^+,\pi^-p)$ cross
section was interpreted as resulting from a contribution of a preexisting
$\Delta^-$ component of the nuclear wave function.  The corresponding data
analysis points to the probability of existence of the virtual $\Delta$
isobars in the above mentioned nuclei at a level of $0.5\div3\%$.

The photo- and electroproduction processes as a test of $\Delta$ admixture in
nuclei were considered in Ref.~\cite{Lipkin}. In Ref.~\cite{Laget} a combined
study of $(e,e'\Delta^{++})$ and $(e,e'\Delta^o)$ reactions on $^{3}$He was
proposed.  The corresponding investigations are being realized at MAMI
(Mainz)~\cite{Kuss} and Jefferson Lab.~\cite{Berman}. Results for the reaction
$^3$He$(e,e'\pi^{\pm})$ obtained at MAMI do not contradict the assumption that
a preformed $\Delta^{++}$ contributes to the reaction yield~\cite{blomq}.  As
for the processes with real photons, only the measurements of
$^{12}$C$(\gamma,\pi^+p)$ and $^{3}$He$(\gamma,\pi^{\pm}p)$ reactions have
been reported in Refs.~\cite{McKenz97} and~\cite{huber}.  The
$^{12}$C$(\gamma,\pi^+p)$ reaction was explored at MAMI in order to study the
mechanisms connected with final state interaction (FSI)~\cite{McKenz97}.  An
estimation of the $\Delta$ admixture in $^{12}$C from these data is difficult
since the experimental cross sections were obtained mainly at small recoil
momenta of the residual nucleus where the background effects are large.  The
probability of the $\Delta NN$ configuration in the $^3$He ground state of
$2.2\pm1.3\%$ was reported in Ref.~\cite{huber}.

The small size of the isobar admixture in nuclei hampers the experimental
observation of the $\Delta$ component. The main difficulty consists in
separating the different background mechanisms leading to the same final
state.  We have solved this problem by measuring the photoproduction cross
section for $\pi^+ p$ pairs in the reaction
\begin{equation}\label{gammaA}
\gamma + ^{12}C \rightarrow  \pi^+p + X\,.
\end{equation}
The measurements were performed in the kinematical region of high momentum
transfer to the residual nuclear system $X$, whereas the background processes
favour the small momentum transfer range.  The use of reactions of the type
(\ref{gammaA}) has two main advantages, outlined also in our previous
paper~\cite{FGK99}. Firstly, charge conservation prevents the creation of the
$\Delta^{++}$ isobar by photons impinging on a single nucleon. Thus, the
interpretation of the experimental results in terms of knocked-out $\Delta$
isobars is simplified. Secondly, since the virtual $\Delta$ isobars are
produced in NN collisions, it is reasonable to expect that the probability of
finding a $\Delta$ isobar in a nucleus is quadratic in the nuclear
density. Therefore, the search for the $\Delta$ components in electromagnetic
reactions is preferred since the photon beam is able to probe the whole
nuclear volume up to the region of the highest densities.  Also because of the
reduced initial state interaction, a more direct test of isobar configurations
can be carried out with electromagnetic probes than with strongly interacting
probes.

Our experimental setup is described in section~2. In section~3 we briefly
summarize the general formalism of the $\Delta$ knock-out model.  Next the
estimation of the background contributions as well as the analysis of the data
are performed.  After the determination of the kinematic region where the
background effects are negligible, we analyse the data under the assumption
that the $\pi^+p$ pairs found in the detectors are the product of the
$\gamma\Delta^{++}\rightarrow\pi^+p$ process which starts from the
$\Delta^{++}$ constituent preexisting in a target nucleus.  This allows us to
determine a value of the $\Delta^{++}$ momentum distribution and the number of
$\Delta$ isobars in $^{12}$C.

The preliminary results of the present experiment have been published 
in Ref.~\cite{pisma}.

\section{Experiment}

Measurements of the differential cross section of the reaction
$^{12}$C$(\gamma,\pi^{+}p)$ were performed at the Tomsk synchrotron at two
electron beam energies, $E_e$ = 420 and 500~MeV, respectively. The electron
beam exhibited a duty cycle of 6\% and had an energy spread of $0.2\%$.  The
electrons hit a 0.4~mm thick Ta-radiator and produced bremsstrahlung photons.
After leaving the vacuum tube of the accelerator, the photon beam was confined
by two lead collimators with an aperture of 6~mm and 10~mm in diameter and
cleaned in a sweeping magnet (see Fig.~1). At the target position the photon
beam had a spot size of 16~mm in diameter and an intensity of about $10^8$
equivalent quanta per second. The shape of the bremsstrahlung spectrum was
described by the Schiff distribution~\cite{Schiff}. The total energy of the
photon flux during the runs was measured by the Gauss-quantameter ($Q$ in
Fig.~1) with an accuracy of $3\%$~\cite{quant}. The photon beam hit the
target at an angle of $45^\circ$.  A carbon plate of a natural isotopic
composition and $4\times 10^{22}$ nuclei/cm$^2$ in thickness was employed as a
target.  Its dimensions were 4~cm in height and 6~cm in width.

As is shown in Fig.~1, the experimental setup includes two coplanar arms
detecting the positive pion and the proton in coincidence.

A strong focusing magnetic spectrometer~\cite{PTE} was set at an
angle~$\theta_\pi$ = 54$^\circ$ with respect to the photon beam in order to
select $\pi^+$ with mean momentum~$p_\pi$ = 181~MeV/c.  Earlier this
spectrometer was used for measurements of a deuteron photodisintegration
asymmetry below pion threshold~\cite{asym} and in measurements of the
$^{12}$C$(\gamma,\pi^{-}p)$ reaction cross section in the second resonance
region~\cite{JaFiz}. The angular and the momentum acceptances were determined
by a Monte Carlo calculation taking into account the beam size, the angle of
the target with respect to the beam, multiple scattering, energy loss, and
pion decay. The solid angle $\Delta\Omega_\pi$ covered $3\cdot 10^{-3}$~sr
(azimuthal and polar angular acceptances were $\Delta\varphi_\pi \approx
6.4^{\circ}$ and $\Delta\theta_\pi \approx 2.0^{\circ}$, respectively) and the
momentum acceptance $\Delta p_\pi/p_\pi$ was $24\%$.  The scintillation
hodoscope~\cite{PR85IFVE} located in the focal plane of the spectrometer
determined the pion momentum with a resolution of $\sigma _{p_\pi}$ = 1.2~\%.
This value includes the contribution due to those muons from pion decay in
flight which reached the telescope.

The proton channel included a $\Delta E-E$ plastic scintillation counter
hodoscope and two auxiliary counters $S_m$ with absorbers intended for
monitoring the proton channel stability. The $E$ detector included three
scintillation counters of the dimensions $10\times10\times50$~cm$^3$, which
were located on top of each other at a distance of 64 cm from the target.  A
single $\Delta E$ counter in front defined the solid angle of the proton
channel to $\Delta\Omega_p$ = 0.26~sr. The average polar angle of the proton
channel with respect to the photon beam and its angular coverage were
$\theta_p$ = $(75\pm19)^\circ$, respectively. The scintillators of the $E$
counters were read out by two photomultiplier tubes FEU-139 at each end, and
the scintillator of the $\Delta E$ counter by two photomultiplier tubes
FEU-30. The absolute energy calibration of the proton channel and the
measurement of the proton counter responses as functions of the particle
coordinate position were performed by using a secondary proton beam selected
by the magnetic spectrometer of the pion channel by positioning the $\Delta
E-E$ counters behind the magnet.  Within the range of the proton energy $T_p =
40\div130$~MeV the analysis of the signals from the photomultipliers of the
$\Delta E$ and $E$ counters allowed to determine the polar angle and the
energy of the detected proton with position independent accuracies better than
$\sigma_\theta=3^{\circ}$ and $\sigma_E=4$~MeV, respectively. The nuclear
interaction of the proton with the scintillators contributed in part to these
values which where determined by the calibration.  The accuracy of the
azimuthal angle~$\phi$ was determined by the counter dimensions to
$\sigma_\phi\sim2^{\circ}$.

For online monitoring of the stability of the proton detector, two narrow
auxiliary scintillators $S_m$ with absorbers have been set up behind the
$\Delta E-E$ hodoscope (see Fig.~1). Simultaneously with the accumulation of
the $\pi^+p$ events we measured individual pulse height spectra of the $\Delta
E$ and $E$ scintillators, which were triggered by the coincident pulses from
the $\Delta E,E$- and $S_m$-counters. Two ranges of the spectra corresponding
to minimal pion and minimal proton energies were identified and
monitored. These energies were determined by the absorber thickness and they
were chosen within the proton channel operating range in such a way that light
responses in the $E$-counters were corresponding to $T_p\approx 52.5$~MeV and
$T_p\approx 77.2$~MeV, respectively.

To reduce cosmic background, the focal plane detectors of the pion channel
were covered with a big-area scintillation counter $S_a$ working in the
anti-coincidence mode.  The final trigger was formed only during the
accelerator pulse, which provided a cosmic background suppression to the level
of less than 2$\%$.  Further data analysis suppressed the cosmic background to
a negligible fraction.

Background measurements were taken using only the pion channel with mean
momentum of 181~MeV/c. The non-target contribution was measured as a ratio of
pion yields with and without the target. The ratio was found to be less than
0.7$\%$.

The measured reaction yields were corrected for the following effects:
\begin{enumerate}
 \item The level of the accidental coincidences depended on the proton energy
       and was estimated from the intensity of the events lying outside the
       time-correlation peak of the $\pi^+p$ coincidences.  This level changed
       from 6$\%$ to $1.5\%$ in the region $T_p = 50\div80$~MeV and remained
       less than $1.5\%$ for $T_p>80$~MeV;
 \item A portion of the particles was lost due to nuclear interaction of the
    protons and pions in the material of the
       target and detector. The associated correction was momentum dependent
       and reached 6$\%$ at maximum;
\item The pion decay correction was found to be 18~$\%$;
 \item The non-target contribution (air and target holder) was 0.7$\%$;
 \item The dead-time due to data recording was 2$\%$.
\end{enumerate}

In the present experiment fifty-three $\pi^+p$ events have been detected in
total. Fig.~2 shows the distribution of these events in the proton emission
angle and the proton energy for the two values of electron beam energy.

\section{Results and discussion}

Before dealing further with the analysis of the obtained results we would like
to outline the general ideas which have governed the calculations for this
investigation. Firstly, we note that the distribution of events shown in
Fig.~2 exhibits a rather pronounced structure formed by the two groups at $T_p
< 80$~MeV and $T_p > 80$~MeV. In the kinematic conditions of our experiment
the larger proton energy $T_p$ implies a larger momentum transferred to the
residual nucleus. Since the isobar configurations are responsible mainly for
high momentum components of the nuclear wave function, a higher recoil
momentum of the residual nuclear system is favored in this case. Keeping this
in mind one may assume that the $\gamma\Delta^{++}\to\pi^+p$ reaction on
preexisting $\Delta$'s results in the right groups of the events in Fig.~2. As
is confirmed by a calculation (described below), the remaining reaction
channels forming the background lead to the events localized in the left part
of the figure. In accordance with the observed structure of the distribution
of events we used two different models for the data analysis. The first one,
hereafter called the $\Delta$ knock-out model, assumes the $\pi^+p$-production
to proceed through the decay of the knocked-out $\Delta^{++}$. The second
model, also called in the literature the Valencia
Model~\cite{Carr,Oset1,Oset2}, was explored to describe the essential part of
background reactions. We outline below the main ingredients of both models and
apply them to the analysis of the data.

\subsection{The $\Delta$ knock-out model}

To interpret the experimental data in the region of high momentum transfer, we
used the approach in which the $\pi^+p$ pair production was considered as a
result of the $\gamma\Delta^{++}\rightarrow\pi^+p$ process on a $\Delta^{++}$
preexisting in the target nucleus.  The differential cross section for the
reaction $^{12}$C$(\gamma,\pi^+p)$ within the $\Delta$ knock-out model was
discussed in our earlier work~\cite{FGK99}. The main ideas are presented here
for convenience. The differential cross section may be written in the
form~\cite{FGK99}
\begin{equation}\label{Mfi}
\frac{d^3\sigma }{dE_p d\Omega _p d\Omega _\pi}=\frac{E_f \, E_p \,p_p\,
p_\pi^3\,  }
{4(2\pi )^5E_\gamma \mid E_f p_\pi^2 - E_\pi\ {\bf p}_\pi\cdot {\bf p}_f \mid}
\, \overline{\left |{M_{fi}}\right | ^2}\,,
\end{equation}
where the subscripts $\gamma$, $\pi$, $p$, and $f$ stand for the photon, pion,
proton, and the final nucleus, respectively. The total energies and momenta of
the participating particles are denoted by $E$ and ${\bf p}$.  The squared
modulus of the reaction amplitude $M_{fi}$ in the laboratory system is
\begin{equation}\label{M2}
\overline{\left |{M_{fi}}\right | ^2}
=\rho _{\Delta^{++}}(p)\, f_{\pi p}\,\,
\overline{\left |{T_{\gamma\Delta\to\pi p}}\right | ^2} \,.
\end{equation}
Here, the subscript $\Delta$ stands for the $\Delta$ isobar,
$T_{\gamma\Delta\to \pi p}$ is the elementary $\gamma\Delta^{++}\rightarrow
\pi^{+}p$ amplitude.  The expression~(\ref{M2}) was obtained employing the
impulse approximation and using the closure relation when summing over all the
states of the undetected residual nuclear system.  The function
$\rho_{\Delta^{++}}(p)$ in (\ref{M2}) represents the $\Delta^{++}$ isobar
momentum distribution in the ground state of the initial nucleus. The factor
$f_{\pi p}$ takes into account the absorption of the produced particles
$\pi^+$ and $p$ inside the nucleus.  It depends on the pion and proton
energies and their opening angle.

The momentum distribution $\rho _{\Delta^{++}}(p)$ obeys the
following normalization condition
\begin{equation}\label{norm}
\int{\rho_{\Delta^{++}}(p)\frac{
d{\bf p}}{(2\pi)^3}}=A\, N_{\Delta ^{++}}^c\,\, ,
\end{equation}
where $N_{\Delta ^{++}}^c$ is the number of $\Delta^{++}$ isobars per nucleon
in the ground state of the $^{12}$C nucleus and $A$~=~12 is its mass number.

The elementary $\gamma \Delta^{++}\rightarrow \pi^{+}p$ amplitude
$T_{\gamma\Delta\to\pi p}$ was obtained within the diagrammatic approach.  The
model was based on the coupling of photons to pions, nucleons and $\Delta$
isobars using effective Lagrangians. This leads to a set of the Feynman
diagrams at the tree level shown in Fig.~3\,(a-e). In the nonrelativistic
limit up to the order of $(p/M)^2$ the elementary amplitude can be written as
(for more details see Ref.~\cite{FGK99})
$$
T_{\gamma\Delta\to\pi p} = i\,
\frac{\sqrt{2}\,f_{\pi N\Delta }}{m_\pi }\frac e{2\,M_p}
\left\{
2{\bf S}^{+}
\cdot\,{\bf p}_{\pi p}
\frac{2\,{\bf p}_\Delta \cdot\,{\mbox{\boldmath$\varepsilon$}}_{\lambda}
\, + i\,\displaystyle \frac{\mu_{\Delta^{++}}}{3}
\,{\mbox{\boldmath$\sigma$}}_{\Delta} \cdot\,[{\bf p_\gamma}\times
{\mbox{\boldmath$\varepsilon$}}_{\lambda} ]}
{E_\Delta +E_\gamma-E_\Delta^{\prime}+i\,\Gamma_\Delta / 2}  \right.
$$
$$
+\,i\,\frac{f_{\pi \Delta \Delta }}{f_{\pi N\Delta }}\,\frac{3\,\mu
_{N\Delta }
{\bf S}^{+}
\cdot\, [{\bf p}_{\gamma \Delta }\times {{\mbox{\boldmath$\varepsilon$}
}_{\lambda}]}}
{E_\Delta -E_\pi
-E_{\Delta}^{\prime\prime }}\,{\mbox{\boldmath$\sigma$} }_\Delta\cdot{\bf
p}_\pi
+ 2\,\frac{{\bf p}_p{ \cdot\,{\mbox
{\boldmath$\varepsilon$}}}_{\lambda}\, +i\,\displaystyle{\frac{\mu
_p}2}\,{\mbox{\boldmath$\sigma$}}\cdot\,[{\bf p_\gamma\times
{\mbox{\boldmath$\varepsilon$}}_{\lambda} ]}}
{E_\Delta-E_\pi -E_p^{\prime }}{\bf S}^{+}\cdot\,{\bf p}_{\pi p}
$$
\begin{equation}\label{amplitude}
\left.
-\, \frac{4\,M_p}{t-m_{\pi}^2}
{\bf S}^{+}\cdot\,({\bf p}_\pi-{\bf p}_\gamma)
\,{\bf p}_\pi\cdot\,{\mbox{\boldmath$\varepsilon$}
}_{\lambda}\,
-\,2\,M_p {\bf S}^{+}\cdot\,
{\mbox{\boldmath$\varepsilon$}}_{\lambda} \right\}.
\end{equation}
Here $\lambda=\pm1$ and ${\mbox{\boldmath$\varepsilon$}}_{\lambda}$ are the
index and vector of the photon polarization.  The variables
$$
E'_p=M_p+({\bf p}_\Delta-{\bf p}_\pi)^2/2M_p,
$$
$$
 E''_{\Delta}=M_{\Delta}+({\bf p}_\Delta-{\bf p}_\pi)^2/2M_{\Delta}
$$
and
$$
E'_{\Delta}=M_{\Delta}+({\bf p}_\Delta+{\bf p}_\gamma)^2/2M_{\Delta}
$$
are the energies of the intermediate nucleon and the $\Delta$. The vectors
$$
{\bf p}_{\pi p}=({\bf p}_\pi M_p-{\bf p}_pE_{\pi})/(M_p+E_{\pi})
$$
and
$$
{\bf p}_{\gamma \Delta}=({\bf p}_\gamma M_{\Delta}-({\bf p}_{\Delta}-{\bf
p}_\pi)E_{\gamma})/(M_{\Delta}+E_{\gamma})
$$
are the relative momenta in the $\pi p$ and $\gamma \Delta $ systems.  In
(\ref{amplitude}), $\bf {S}^{+}$ is the transition operator between the states
with the spin 1/2 and 3/2 and $\mbox{\boldmath$\sigma$}_\Delta$ is the analog
of the Pauli spin matrix for the spin $\frac32$ object (full expressions of
$\bf S$ and $\mbox{\boldmath$\sigma$}_\Delta$ are given e.g.\ in
Ref.~\cite{Aren78}).  For the hadronic coupling constants we employ $f_{\pi
N\Delta}^2/4\pi$\,= 0.37 from the decay $\Delta \rightarrow \pi N$ and
$f_{\pi\Delta\Delta}\,=\,4/5 f_{\pi NN}$, as predicted by the trivial quark
model.  The magnetic moments used in this calculation are $\mu _{p}$\,= 2.79,
$\mu_{\Delta^{++}}$\,= 4.52~\cite{Wall95} and
$\mu_{N\Delta}$\,= 3.24~\cite{Carl} in terms of nuclear magnetons.

\subsection{The Valencia Model}

The Valencia model~\cite{Carr,Oset1,Oset2} is well suited to estimate the
background contributions to the direct $\Delta$ knock-out reaction under
investigation.

It has already been applied in several studies of similar
reactions~\cite{McKenz97,Oset1,Lamp,Harty}.  This model incorporates all major
photoabsorption mechanisms as one-, two- and three-nucleon absorption as well
as single-pion production. These photoabsorption mechanisms are treated
microscopically in a quantum-mechanically correct way taking into account the
relevant photon-nucleon interaction amplitudes. Within the local density
approximation the nuclear matter results are transferred to finite nuclei. The
nuclear many-body problem is simplified by using a Fermi gas model. The
particles which are produced in the photoabsorption process are then
propagating independently through the target nucleus. They are traced in a
phenomenological way considering the most important N-N and $\pi$-N reaction
channels until the primary and/or secondary hadrons leave the target
nucleus. One of the main assumptions of this model is the treatment of the
knocked out particles on their path through the medium as on-shell particles.
The angular distributions are included in a phenomenological manner.  With
these ingredients for each event, the final momentum of each hadron is the
result of a Monte Carlo type calculation.

The Valencia model describes total cross sections or dominant reaction
channels in the $\Delta$ resonance region, e.g. as ($\gamma$,np) or
($\gamma$,$\pi$,n), quite well (Refs.~\cite{Lamp,Harty}).  However, for some
of the less important reaction channels sizeable deviations can be found as
these are sensitive mostly to final state interactions. For the particular
kinematics of Ref.~\cite{Harty} a 50\% overestimation of the
($\gamma$,p$\pi^\pm$) cross sections was found for the photon energy range of
$E_\gamma=400\div600$~MeV.

In order to investigate the sensitivity within the present kinematics some
parameters have been varied.  In Fig.~4 we present the upper and lower
boundaries for the cross section given by the Valencia Model at the photon
energy $E_\gamma$ = 450~MeV.  In particular, to account for nucleon-nucleon
correlations the nucleon wave functions are modified by a Jastrow type
correlation $j_\circ(q_cr)$.  This has a direct effect on the momentum
distributions and thus at the relative strength of competing reaction
channels.  The many-body cut-off $q_c$, whose standard value involved in the
model is 783~MeV/c, has been varied in the region $650\div900$ ~MeV/c.  The
important observation of these trial calculations is the proton energy
spectrum which changes shape and strength, however, above $T_p = 80$~MeV the
background contribution remains very small.

\subsection{Data analysis for cross sections}

With the aim of determining the kinematic region where the reaction mechanisms
caused by non-nucleon degrees of freedom may dominate, the analysis was
carried out for the contributions from the background $\pi^+ p$ pair
production mechanisms. For this purpose we employed the Valencia model.

The calculations were performed for three photon energies, namely $E_{\gamma}$
= 350, 400 and 450~MeV.  For the cut-off momentum we have used $q_c$ =
783~MeV/c adopted as a standard value of this parameter in this model.  The
differential cross sections as a function of the proton kinetic energy~$T_p$
for $E_{\gamma}$ = 400 and 450~MeV have been averaged over the acceptances of
the other observables and are shown in the histograms of Fig.~5.  No result
for the lowest photon energy is presented, because the corresponding cross
section is negligible. According to the Valencia model, the $\pi^+ p$ yield at
our experimental conditions results mainly from the sequential processes:
$\pi^o p$ and $\pi^+ n$ pair production followed by charge-exchange
rescattering of the $\pi^o$-mesons to $\pi^+$-mesons and the neutrons to
protons. As one can see in Fig.~5, the maximum of the background reactions is
located at the small proton energies.  This is due to the fact that the
quasi-free $\pi^{+}n$ and $\pi^o$p production dominates in the region of small
momenta of the residual nuclear system, which in the kinematic conditions of
our experiment corresponds to small proton energies. This is corroborated by
Fig.~6, where the residual nucleus momentum~$p_f$ is presented as a function
of the proton energy~$T_p$ for various values of the proton angle~$\theta _p$.

Note that all estimates for the competing reaction mechanisms predict
contributions dominantly for the proton energies less than 80~MeV.  The
estimate of the background reactions according to Valencia model is 4\% for
$T_p>80$~MeV. Also we would like to note that this conclusion is independent
from the values chosen for the parameters of the Valencia model. Fig.~4 shows
that the contribution from the background reactions remains rather insensitive
to their variation in the region $T_p>$80~MeV.

To analyze the data for $E_e$ = 500~MeV, we took into account the background
from the double pion production in the $^{12}$C$(\gamma,p\pi^+\pi^-)^{11}$B
reaction, which was absent at $E_e$ = 420~MeV. This photoabsorption mechanism
is not included in the Valencia model.  An estimate of the double pion
production background has been done within the framework of the quasi-free
approximation, whose formalism is presented in detail in Ref.~\cite{pipi}. For
proton energies $T_p>$ 80~MeV the calculation shows that the net effect from
this process is less than $3\%$ at $E_e$ = 500~MeV (see Fig.~5).

As can be seen from Fig.~2, a considerable reaction yield is observed in the
range $T_p\geq 80$~MeV. Therefore, we interpret these events to be caused by
the binary process $\gamma\Delta^{++}\rightarrow\pi^+p$ on a preexisting
$\Delta^{++}$.

A further support of this idea is provided by the analysis of the $\pi^+p$
invariant mass distribution presented in Fig.~7. Despite the statistics one
sees that the events from the region $T_p\geq80$~MeV fall within the range of
the $\Delta(1232)$, which confirms once again our estimate about the reaction
mechanism in this energy region.  Due to these reasons, we have selected for
further analysis only those events, which lie in the range $T_p\geq 80$~MeV
(right of the dashed line in Fig. 2).  Even when combining both runs there are
only 13 events left.

As a result, we determined the differential yield ${d^4Y}/{dE_\pi d\Omega_\pi
dE_pd\Omega_p}$, which is related to the measured quantities and to the
differential cross section by the equations
\begin{equation}\label{Y}
\frac{d^4Y}{dE_\pi d\Omega_\pi dE_pd \Omega_p}=
  \frac{N_{\pi p}E_{\gamma max}}{\Delta E_\pi \, \Delta \Omega_\pi \,
                    \Delta E_p \, \Delta \Omega_p \, W_\gamma \, t}\, ,
\end{equation}
\begin{equation}\label{Ex}
\frac{d^4Y}{dE_\pi d\Omega_\pi dE_pd\Omega_p}=
  \frac{d^3\sigma}{ dE_pd\Omega_pd\Omega_\pi} \, f(E_\gamma) \, \Biggl|
  \frac{\partial E_\gamma}{\partial E_\pi} \Biggr | \,\,,
\end{equation}
\noindent
where $N_{\pi p}$ is the number of events in the phase space determined by the
intervals $\Delta E_\pi$, $\Delta E_p$, $\Delta\Omega_\pi$, and
$\Delta\Omega_p $. The target thickness is denoted by $t$ (in nuclei/cm$^2$).
The Jacobian
\begin{equation}
\frac{\partial E_\gamma}{\partial E_\pi} =
   \frac{E_f \, p_\pi^2-E_\pi {\bf p}_\pi {\bf p}_f}
        {E_\gamma E_f -  {\bf p}_\gamma {\bf p}_f} \,
   \frac{E_\gamma}{p_\pi^2}
\end{equation}
\noindent
connects the intervals $\Delta E_\pi$ and $\Delta E_\gamma$.  In equation
(\ref{Y}) $W_\gamma$ is the total energy of the photon flux, and $E_{\gamma
max}=E_e -m_ec^2$ stands for the endpoint energy of the bremsstrahlung
spectrum $ f(E_\gamma) $, which is normalized as
\begin{equation}
 \int_0^{E_{\gamma max}} f(E_\gamma)\,E_\gamma\ dE_\gamma= E_{\gamma max}\ .
\end{equation}
The number of photons in the interval $\Delta E_\gamma$ is determined
by the relation
\begin{equation}
N_\gamma (E_\gamma)\, \Delta E_\gamma = \frac{W_\gamma}{E_{\gamma max}}
\, f(E_\gamma)\, \Delta E_\gamma \ .
\end{equation}

The quantities not measured in the reaction, $E_\gamma$ and $p_f$, were
determined by solving a set of kinematic equations
$$ E_\gamma + M_{C^{12}} = E_\pi + E_p + E_f $$
\begin{equation}
{\bf p}_\gamma = {\bf p}_\pi +{\bf p}_p + {\bf p}_f
\end{equation}
\noindent
under the assumption that the residual nucleus $^{11}$Be is in its ground
state.  In order to estimate the error connected with this assumption, we have
repeated the calculation considering the residual nucleus to be excited with
the energy $E_x$ = 20~MeV. The excitation energy was taken as an average of
the excitation spectrum obtained from the $^{12}$C$(e,e'p)$ data~\cite{eep}.
This assumption leads to the difference in $N_\gamma (E_\gamma)\, \Delta
E_\gamma$ and correspondingly in the value of the cross section of about 5$\%$
as compared with the choice $E_x$ = 0~MeV.  This value was also taken into
account in the systematical experimental error.

The cross section, obtained from the reaction yields at the two electron
energies $E_e$ = 420 and $E_e$ = 500~MeV, was averaged over the intervals:
\begin{eqnarray}\label{Sp}
T_p      & = & 80   \div   120~\mbox{MeV};\\
T_\pi    & = & 71.5 \div 106.5~\mbox{MeV}; \nonumber\\
\theta_p & = & 56   \div    94^{\circ}\,. \nonumber
\end{eqnarray}
The average photon energy for the events lying in the kinematic region under
consideration was 355~MeV.

Within the procedure described above, we have obtained the following results
for the runs with electron beam energy $E_e$ = 500~MeV and $E_e$ = 420~MeV,
\begin{equation}\label{dsigma}
\frac{d^3\sigma}{ dE_pd\Omega_pd\Omega_\pi}
\left[ \frac{\mbox{nb}}{\mbox{MeV~sr}^2}\right] = \left\{
\begin{array}{ll}
 7.5 \pm 3.4 \,\, _{-\,\,0.4}^{+\,\,0.7} \  &
                                  \mbox{run at}\, E_e=500\, \mbox{MeV} \\
& \\
10.7 \pm 3.8 \,\, _{-\,\,0.5}^{+\,\,0.8} \  &
                                  \mbox{run at}\, E_e=420\, \mbox{MeV} \\
& \\
 8.9 \pm 2.6 \,\, _{-\,\,0.5}^{+\,\,0.8} \  & \mbox{sum} \, ,
\end{array} \right.
\end{equation}
where statistical and systematical errors are given.  The third value in
(\ref{dsigma}) is the weighted average of the individual results. The
systematical error includes the uncertainties of the corrections mentioned in
the listing of section 2 and the measurement of the photon flux energy.

The experimental cross section is shown in Fig.~5 by the single point where
the vertical bar represents the total uncertainty as quadratic combination of
both errors. The horizontal bar is the standard deviation obtained for $T_p$
under the assumption that the events are uniformly distributed over the
averaging interval (see Fig.~2).

\subsection{The $\Delta$ momentum distribution}

The knock-out reactions are well known to be a powerful tool to study the
momentum distribution of nucleons in nuclei. If we treat the $\Delta$ isobar
essentially as an ordinary particle carrying its energy and momentum, it is
reasonable to speculate about the $\Delta$ momentum distribution inside the
nucleus. The basis of our $\Delta$ knock-out model supports this suggestion as
can be seen from expressions~(\ref{Mfi})~to~(\ref{norm}) where the
$\Delta^{++}$ momentum distribution $\rho_{\Delta^{++}}(p)$ in $^{12}$C is
used. One of the aims of the present experiment was to measure the function
$\rho_{\Delta^{++}}$ within a certain range of ${p}$.  The average of the
$\Delta^{++}$ momentum in the present experiment amounts to
$\overline{p}=300$~MeV/c, and the standard deviation of the momentum
distribution is 49~MeV/c. This value $\overline{p}$ was determined in the
frame of the quasi-free approximation with
\begin{equation}
{\bf p}=-{\bf p_f}\,,
\end{equation}
where ${\bf p}$ is the momentum of the $\Delta^{++}$ isobar and ${\bf p_f}$ is
the momentum of the residual nucleus, estimated on the basis of the
experimental data. Using the expressions (\ref{Mfi})~to~(\ref{norm}) and
averaging the cross section over the intervals defined by (\ref{Sp}), the
value $\rho _{\Delta^{++}}(\overline{p})$ was determined by fitting our
calculation to the weighted average of the data obtained in the two runs (last
value in (\ref{dsigma})).  This approach yields for the mean momentum
$\overline{p}$~=~300~MeV/c
\begin{equation}\label{Exrho}
\rho _{\Delta^{++}}(\overline{p})=
0.17\pm 0.05 \pm 0.02\,\, \mbox{fm}^3\,,
\end{equation}
where the latter two numbers show the statistical and systematic uncertainties
retained from the cross sections. In addition, the systematic uncertainty
includes the estimate of the background reactions according to the Valencia
model and double pion photoproduction.

In the calculation of (\ref{Exrho}) the suppression factor $f_{\pi p}$ in
(\ref{M2}) was calculated within the optical model using the eikonal
approximation.  We employed the proton-nucleus optical potential obtained in
Ref.~\cite{lage} from a phenomenological study of proton scattering off
nuclei.  This optical potential has allowed to explain an effect of
interaction of the proton with the residual nucleus in the reaction
$^{12}$C$(\gamma, \pi^- p)$ in the range of the $\Delta(1232)$~\cite{Fsi}.
The pion wave function was distorted by an optical potential calculated in
Ref.~\cite{Futami} according to Ref.~\cite{FrankGW}. The authors~\cite{Futami}
have described satisfactorily the inclusive charged pion spectra, produced on
carbon and copper in the kinematical region of quasi-free pion
photoproduction~\cite{Baba:78,Baba:79}. The overall quality of these
calculations permits an estimate of the precision of $f_{\pi p}$ to 26\%.

It should be kept in mind that our quantitative conclusion also depends on the
calculation methods of some nuclear effects and on the constants used in the
$\Delta$ knock-out model. One of the uncertainties is connected with the
$\Delta^{++}$ magnetic moment.  We have chosen $\mu_{\Delta^{++}}=4.52$
nuclear magnetons according to the experimental value given in
Ref.~\cite{Wall95}. It is necessary to remark that many modern theoretical
calculations give the $\Delta^{++}$ magnetic moment close to this experimental
value (see Table~\ref{tab:comp}). We attach a model uncertainty of 18\% with
this factor.

\begin{table}[h]
\begin{center}
\caption{\label{tab:comp}
Comparison of the magnetic moment $\mu_{\Delta^{++}}$
predicted by some resent developments.
}
\vspace{5mm}
\begin{tabular}{|c|l|} \hline
$\mu_{\Delta^{++}}$ in nuclear magnetons
& \ \ \ \ \ \ \ \ \ \ \ \ Comments \\ \hline
$4.52\pm 0.5\pm 0.45$ & SIN data \cite{Wall95}. \\ \hline
$4.13\pm 1.3$ & QCD sum rules \cite{Lee98}. \\ \hline
$4.4\pm 0.8$ &  light cone QCD sum rules \cite{ali2000}. \\ \hline
$4.91\pm 0.6$ & lattice QCD \cite{Lein92}. \\ \hline
$4.0\pm 0.4$ &  chiral perturbation theory \cite{Butl94}. \\ \hline
$4.73$ & chiral quark-soliton model \cite{Kim98}. \\ \hline
\end{tabular}
\end{center}
\end{table}
Our estimates show that the model error introduced by the uncertainty in
magnetic moment of the $\Delta^{++}$ isobar and in the final state interaction
totals to 0.06~fm$^3$.

Unfortunately, we cannot compare our result neither with theory nor with
experiment because of the lack of literature data on the momentum distribution
$\rho _{\Delta^{++}}$ of the $^{12}$C nucleus.  We hope that results of our
experiment will be a stimulus for further detailed studies of the isobar
configurations in nuclei.

\subsection{The number of preformed $\Delta$'s}

In those works, where the isobar configurations in nuclei are studied, the
main attention is concentrated on the estimate of the full number $N_\Delta$
of the $\Delta$ isobars per nucleon in the ground state of the nucleus.  Since
only a small region of the momentum $p$ is covered in our experiment, in order
to estimate $N_\Delta$, it is necessary to know the form of the function $\rho
_{\Delta^{++}}(p)$.  Unfortunately, in the literature the functions $\rho
_{\Delta^{++}}(p)$ are given only for the lightest nuclei such as
deuteron~\cite{Anastas} and $^3$He~\cite{Strueve87} or for infinite nuclear
matter~\cite{Cenni89}.  In the present paper we use the results of the
work~\cite{Cenni89}, where the $\Delta$ momentum distribution in the nuclear
matter was evaluated within Random Phase Approximation.  The free parameters
used in the model~\cite{Cenni89} are the cut-off parameter $\Lambda$, the
coupling constant $f_{\pi N \Delta}$, the Landau-Migdal parameter $g'$, and
the correlation parameter $q_c$. Reasonable variations in these parameters
lead to rather substantial changes of $N_{\Delta}^m$ in the range from 5.66\%
up to 15.89\%.  In the present calculation we have used the occupation number
$n_\Delta^m(p)$ predicted by the choice $\Lambda$ = 1300~MeV/c, $f^2_{\pi N
\Delta}$/4$\pi$ = 0.32, $g'$ = 0.7, $q_c$ = 800~MeV/c recommended in
Ref.~\cite{Cenni89} as the most reasonable one. It gives $N_\Delta^m(p)$ =
6.66~\%, which is in close agreement with the result obtained in
Ref.~\cite{Anastas}.
         
Before making quantitative conclusions we would like to demonstrate that our
idea about the reaction mechanism, i.e. the knocking-out of the preformed
isobars, agrees qualitatively with the results of measurements in the energy
region considered.  The ($\pi^+p$) events from the reaction
$^{12}$C$(\gamma,\pi^{+}p)$ are distributed in six-dimension phase space.
From the possible distributions the invariant mass distribution may
intuitively be assumed to be the most sensitive to the reaction
dynamics. Keeping this in mind, we present in Fig.~7 (solid line) the $\pi^+p$
invariant mass distribution, predicted by the $\Delta$ knock-out model with
the $\Delta$ momentum distribution taken from Ref.~\cite{Cenni89}.  One
can see that the computational results do not contradict the experimental
data.  We consider this fact as indirect evidence of the validity of our
theoretical basis.

Turning to the estimation of the number $N_\Delta$, we used the following
method.  The momentum distribution $\rho _{\Delta^{++}}$ is connected with the
occupation number $n_{\Delta^{++}}^c(p)$ of $^{12}$C by the relation
\begin{equation}\label{rho}
\rho _{\Delta^{++}}(p)=n_S \, \frac 43 \pi
R^{3}\ n_{\Delta^{++}}^c(p),
\end{equation}
where $R=3.2$~fm is the square-well radius for $^{12}$C. The factor $n_S$ = 4
is the number of the $\Delta^{++}$ spin states.  We assume that the momentum
distribution $\rho _{\Delta^{++}}(p)$ of the $\Delta(1232)$ isobars in
$^{12}$C is proportional to that for nuclear matter.  From the
normalization condition~(4) and the analogous condition for nuclear matter we
obtain
\begin{equation}\label{ON}
n_{\Delta^{++}}^c(p)=
\frac{\rho _{N}^{c}}{\rho_{N}^{m}}\,
\frac{N_{\Delta^{++}}^c}{N_{\Delta^{++}}^m}\,
n_\Delta^m(p).
\end{equation}
\noindent
Here $\rho_{N}^{c}$ = 0.087~fm$^{-3}$ and $\rho_{N}^{m}$ = 0.17~fm$^{-3}$ are
the nucleon densities in $^{12}$C and in nuclear matter,
respectively~\cite{Bethe}; $N_{\Delta ^{++}}^m$ represents the number of
$\Delta^{++}$ isobars per nucleon in nuclear matter.

Within $SU(2)$-symmetry the total number of all $\Delta$ isobars per nucleon
in nuclear matter $N_\Delta^m$ is given by
\begin{equation}\label{m}
N_\Delta^m = 4\ N_{\Delta ^{++}}^m.
\end{equation}
The corresponding relation for the total number of the $\Delta$ isobars per
nucleon in $^{12}$C can be written as
\begin{equation}\label{c}
N_\Delta^c = \frac{64}{15}\ N_{\Delta ^{++}}^c.
\end{equation}

We have extracted $N_\Delta^c$ according to formulae
(\ref{Mfi})~to~(\ref{norm}) from the experimental cross section
(\ref{dsigma}) taken as a weighted average of the results obtained in two runs
(latter value in (\ref{dsigma})). Using also $N^m_\Delta=6.66\%$ from
Ref.~\cite{Cenni89} and expressions for $\rho_{\Delta^{++}}(p)$
(\ref{rho})~to~(\ref{c}), we have obtained
\begin{equation}\label{20}
N_\Delta^c=0.017\pm 0.005 \pm 0.002\,\, \Delta\ \mbox{isobars per nucleon}\,,
\end{equation}
where the errors were estimated by the same way as in expression
(\ref{Exrho}). This value is slightly smaller than the result obtained by
other authors~\cite{Morr98}.  The $\Delta$ knock-out model error, estimated as
described in section 3.4, translates to the uncertainty in $N_\Delta^c$ of
about 0.005~$\Delta$ isobars per nucleon.

Finally, we would like to make a comment about the present experimental study
of the $\Delta$-isobars in nuclei.  Using the above parametrization of the
momentum distribution $\rho _{\Delta^{++}}(p)$ with the value $N_{\Delta}$
(\ref{20}), we have calculated the cross section (\ref{Mfi}) as a function of
the proton kinetic energy $T_{p}$.  In Fig.~5 we compare the $\Delta$
knock-out cross section with that for the background reactions and with our
data.  All results are averaged over the intervals given by (\ref{Sp}). One
can see that the maximum of the $\Delta$ knock-out cross section is strongly
shifted to higher proton energies and, hence, is well-defined
kinematically. The experimental cross section at the proton kinetic energy
$T_p >$ 80~MeV is almost two orders of magnitude larger than the theoretical
estimates obtained within the models which allows for only the nucleon degrees
of freedom in the $^{12}$C nucleus. This fact, being particularly important
for low-intensity signals, is one of the main advantages of the method used in
this work.

\section{Conclusions}

The differential cross section for the $^{12}$C$(\gamma,\pi^{+}p)$ reaction
has been measured in the $\Delta(1232)$ resonance region at high recoil
momenta of the residual nuclear system.  By using the $\Delta$ knock-out model
and the Valencia model we have determined the kinematical region where the
background reactions are of minor importance, and the selected $\pi^+p$ pairs
are mainly due to the direct interaction of the photons with the preexisting
$\Delta$ isobars.  Therefore we interpret the events, which lie in this
region, as direct evidence for the $\Delta^{++}$ constituent in the target
nucleus.  Within the $\Delta$ knock-out model the estimate for the
$\Delta^{++}$ momentum distribution
$\rho_{\Delta^{++}}(\overline{p})$=0.17~fm$^3$ for the average momentum
$\overline{p}=300$~MeV/c and the number of $\Delta$ isobars in $^{12}$C
$N_\Delta=0.017$ were obtained.  The derived value for $N_\Delta$ is in
general agreement with the results obtained by other authors for p-shell
nuclei which are mainly in the range of 0.5$\div$3.0\%.

\section*{Acknowledgments}
This work was supported by the Russian Foundation for Basic Research
under the Contracts No. 96-02-16742, No. 97-02-17765, and No. 99-02-16964.

\newpage
{\bf Figure captions}

\vspace{10mm}

Fig.~1. Layout of the experimental setup: (R) radiator, (C) lead collimators,
(SM) sweeping magnet, (VT) photon beam vacuum
tube, (T) target, (Q) Gauss quantameter, (M) analyzing magnet,
(S) scintillation counters, (H) hodoscope, (A) absorbers.
\vspace{5mm}

Fig.~2. The proton emission angle $\theta_p$ is plotted versus the proton
kinetic energy $T_p$ for the final 53 $\pi^+ + p$ coincident events separately
for the two electron beam energies.  The vertical dashed line at $T_p$=80~MeV
separates the accepted events (solid squares) from those which are mainly due
to background (crosses).  The arrow at $T_p$=50~MeV indicates the detection
threshold.  The horizontal dashed lines show the angular range of proton
detection. The vertical dashed line is the cut applied to the proton energy.
\vspace{5mm}

Fig.~3. Diagrams for the $\gamma\Delta^{++}\rightarrow\pi^+p$ amplitude used in
the present calculation: (a) s- channel term, (b) and (c) u-channel terms, (d)
pion pole term, (e) seagull term.
\vspace{5mm}

Fig.~4. Valencia Model calculation for the differential cross section of the 
reaction $^{12}C(\gamma,\pi^+p)$ at $E_\gamma=450$~MeV. The lines 
present the boundaries of the results obtained with different values 
of the model parameters. The cut-off momentum $q_c$ has been varied 
in the range $q=650\div900$~MeV/c.
\vspace{5mm}

Fig.~5. Differential cross section for the $^{12}$C$(\gamma,\pi^{+}p)$
reaction averaged over the intervals $\theta_{p}=(56\div94)^\circ$ and
$T_\pi=(71.5\div106.5)$~MeV as a function of the kinetic energy $T_{p}$: the
results of the Valencia model for single pion production at $E_\gamma$=400 and
450~MeV are shown by the solid and dotted histograms, respectively; the
dot-and-dash curve corresponds to the quasi-free double pion photoproduction;
the solid line corresponds to the $\Delta$ knock-out model; and the dot
represents the experimental cross section with its combined statistical and
systematic uncertainties.
\vspace{5mm}

Fig.~6. Momentum~$p_f$ of the residual nucleus as a function of the kinetic
energy~$T_{p}$ for $\theta_{\pi}=54^{o}$ and $T_{\pi}=90$~MeV.
\vspace{5mm}

Fig.~7. Invariant mass distribution of the $\pi^+p$ events for proton energies
$T_p=80\div120$~MeV. The solid curve is the result of the $\Delta$ knock-out
model calculation.

\newpage
\begin{figure}[h]
\unitlength=1cm
\centering
\begin{picture}(16,10)
\epsfig{file=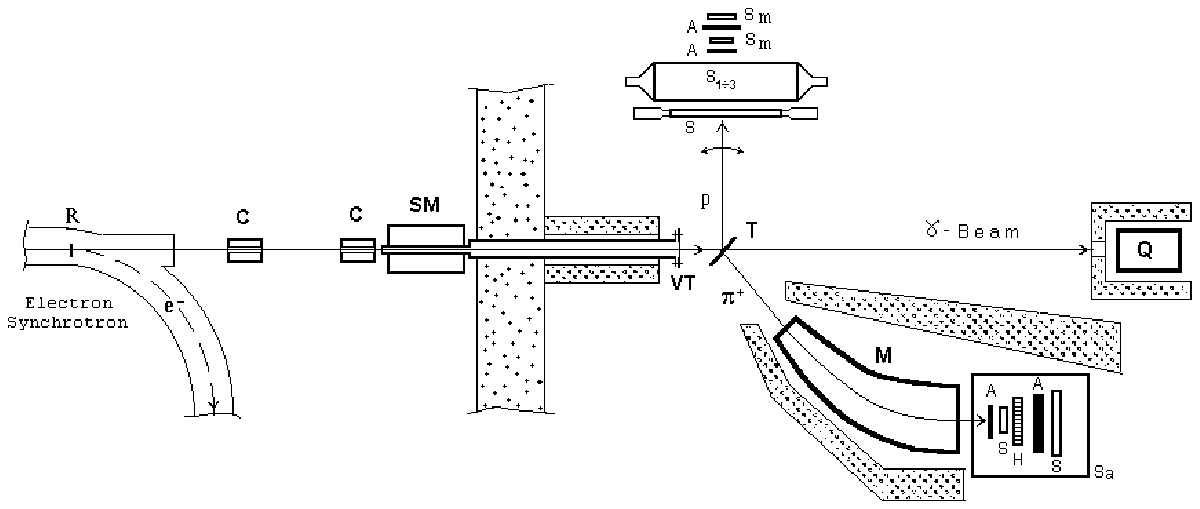,width=.75\textwidth}
\end{picture}
\vspace*{7mm}
\caption{}
\end{figure}

\newpage
\begin{figure}
\unitlength=1cm
\centering
\begin{picture}(10,10) (2,0)
\epsfig{file=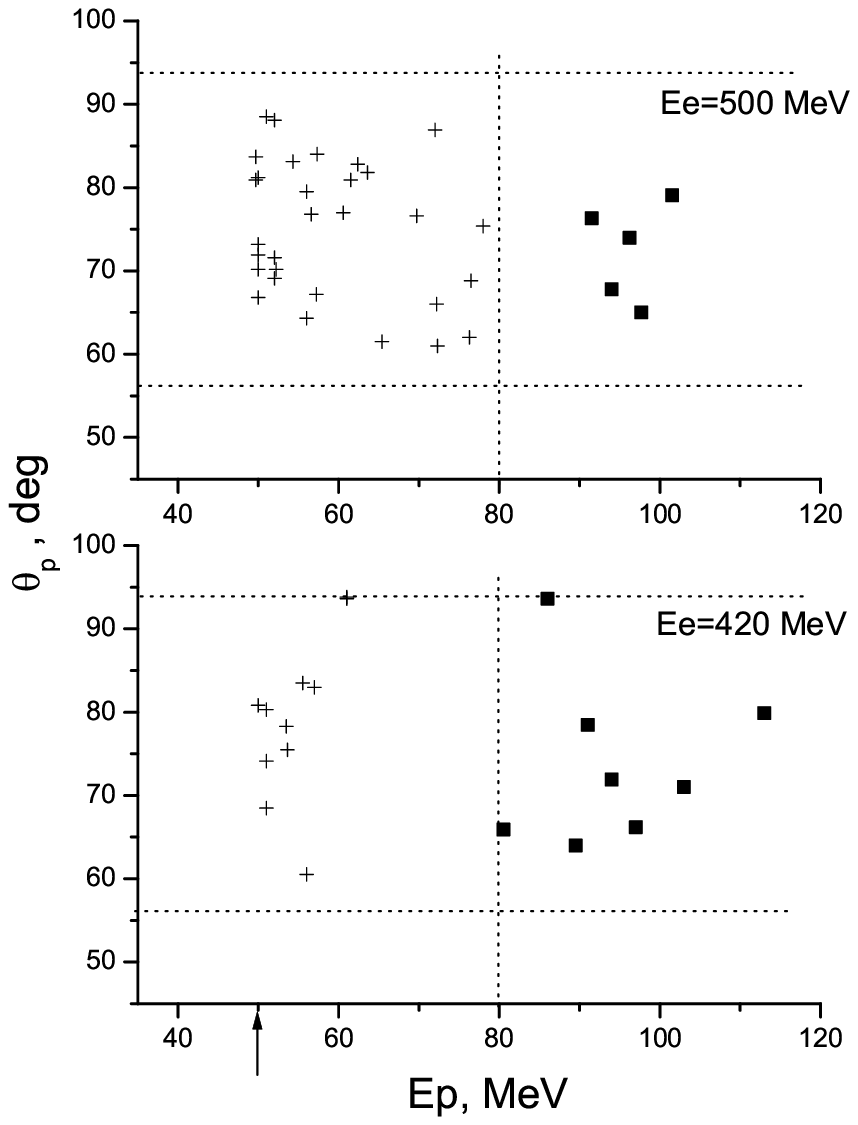,width=.75\textwidth}
\end{picture}
\caption{}
\end{figure}

\newpage
\begin{figure}
\unitlength=1cm
\centering
\begin{picture}(20,20) (0,-4)
\epsfig{file=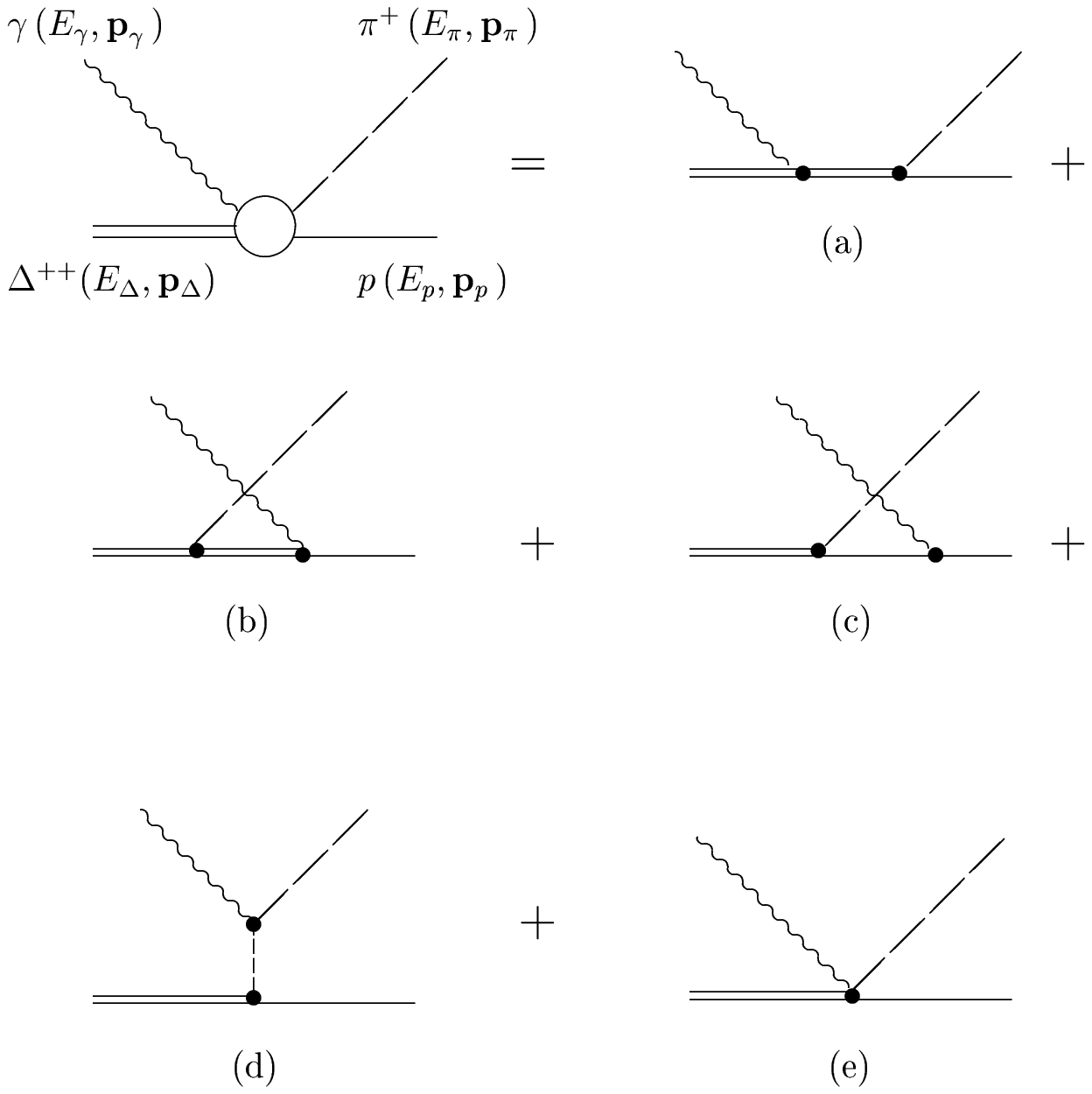,width=.75\textwidth}
\end{picture}
\caption{}
\end{figure}

\newpage
\begin{figure}
\unitlength=1cm
\centering
\begin{picture}(10,10) (2,0)
\epsfig{file=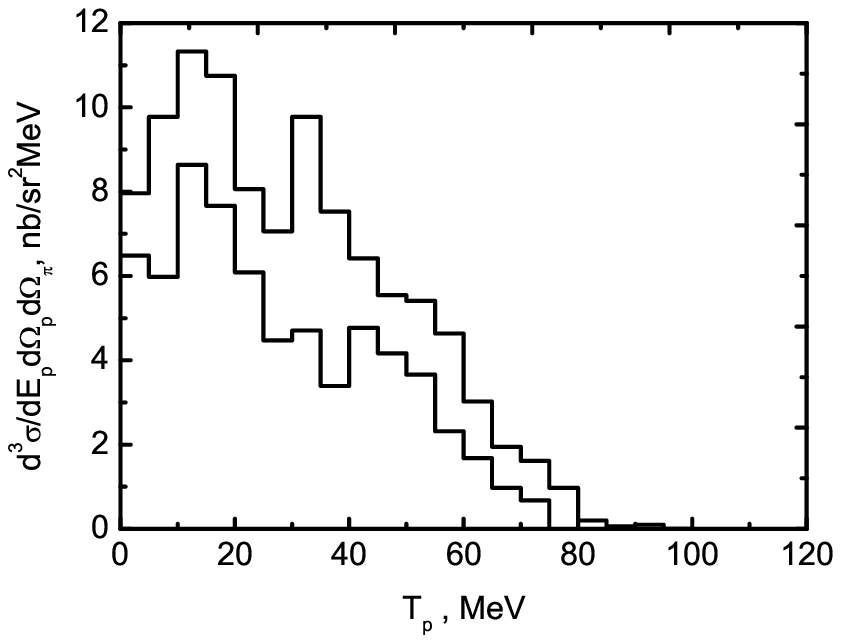,width=.75\textwidth}
\end{picture}
\caption{}
\end{figure}

\newpage
\begin{figure}
\unitlength=1cm
\centering
\begin{picture}(18,13)
\epsfig{file=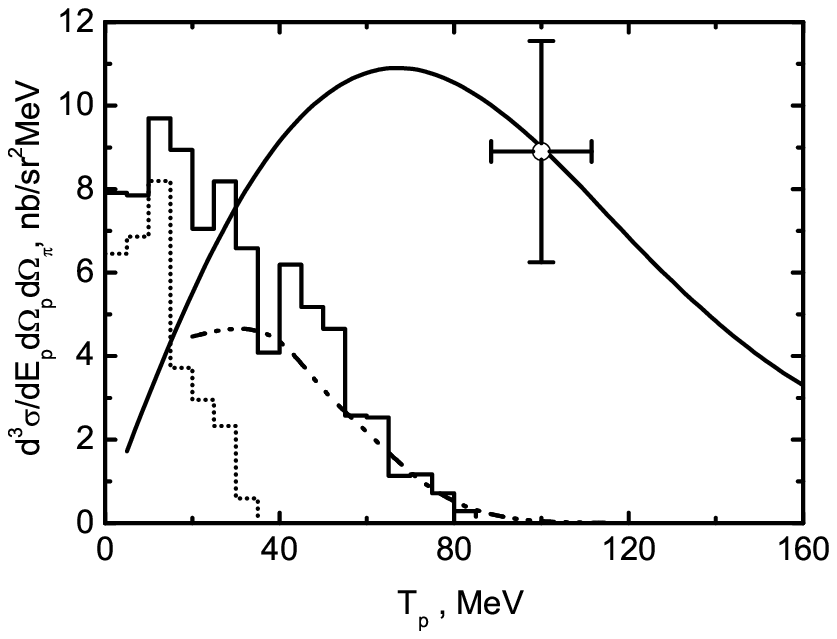,width=.75\textwidth}
\end{picture}
\caption{}
\end{figure}

\newpage
\begin{figure}
\unitlength=1cm
\centering
\begin{picture}(15,12)
\epsfig{file=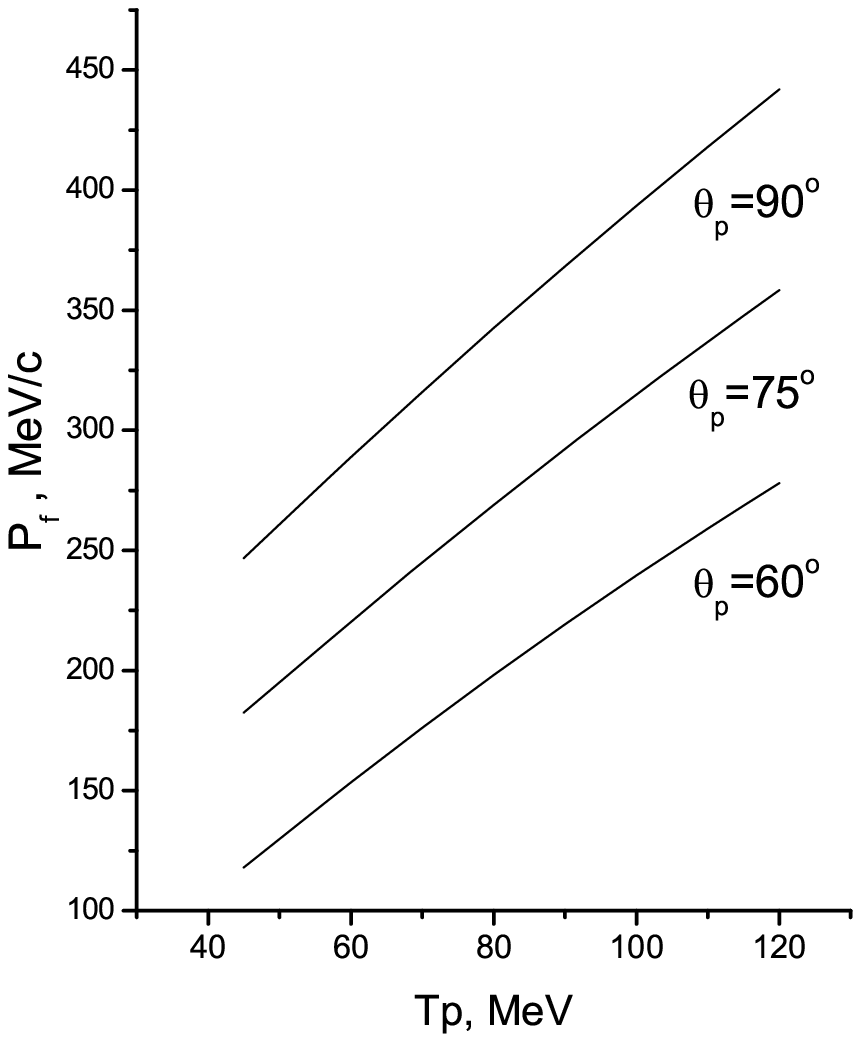,width=.75\textwidth}
\end{picture}
\caption{}
\end{figure}

\newpage
\begin{figure}
\unitlength=1cm
\centering
\begin{picture}(18,15)
\epsfig{file=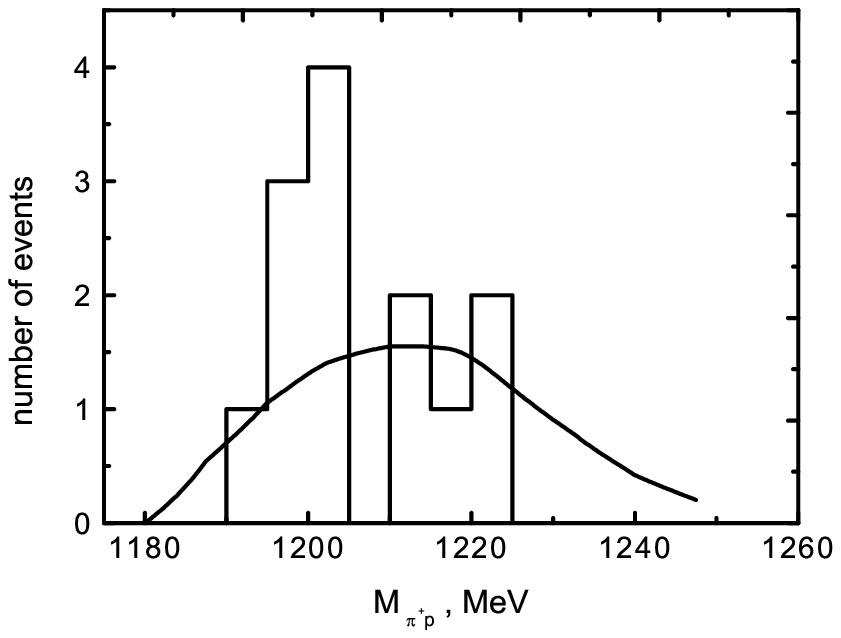,width=.75\textwidth}
\end{picture}
\caption{}
\end{figure}
\end{document}